\documentclass[12pt,preprint]{aastex}

\begin{document}

\title{WISEP J180026.60+013453.1: A Nearby Late L Dwarf Near the Galactic Plane}

\author{John E.\ Gizis}
\affil{Department of Physics and Astronomy, University of Delaware, 
Newark, DE 19716, USA}

\author{Adam J.\ Burgasser}
\affil{Center for Astrophysics and Space Science, University of California San Diego, La Jolla, CA 92093, USA}

\author{Jacqueline K.\ Faherty\altaffilmark{1}}
\affil{Department of Astrophysics, American Museum of Natural History, Central Park West at 79th Street, New York, NY 10034, USA}

\author{Philip J.\ Castro}
\affil{Department of Physics and Astronomy, University of Delaware, 
Newark, DE 19716,USA}

\author{Michael M.\ Shara\altaffilmark{1}}
\affil{Department of Astrophysics, American Museum of Natural History, Central Park West at 79th Street, New York, NY 10034, USA}

\altaffiltext{1}{Visiting Astronomer at the Infrared Telescope Facility, which is operated by the University of Hawaii under Cooperative Agreement no. NNX-08AE38A with the National Aeronautics and Space Administration, Science Mission Directorate, Planetary Astronomy Program.}

\begin{abstract}
We report a nearby L7.5 dwarf discovered using the Preliminary Data Release of the Wide-field Infrared Survey Explorer (WISE) and the Two Micron All-Sky Survey (2MASS).  WISEP J180026.60+013453.1 has a motion of 0.42 arcsec yr$^{-1}$ and an estimated distance of $8.8 \pm 1.0$ pc. With this distance, it currently ranks as the sixth closest known L dwarf, although a trigonometric parallax is needed to confirm this distance. It was previously overlooked because it lies near the Galactic Plane ($b=12^\circ$).  As a relatively bright and nearby late L dwarf with normal near-infrared colors, W1800+0134 will serve as a benchmark for studies of cloud-related phenomena in cool substellar atmospheres.
\end{abstract}


\section{Introduction\label{intro}}

One of the main scientific goals of the Wide-field Infrared Survey Explorer (WISE) \citep{2010AJ....140.1868W} is to identify the brown dwarfs and low-mass stars closest to the Sun. WISE has now surveyed the entire sky in four mid-infrared filters, with $\sim 58\%$ of the sky included in the April 2011 Preliminary Data Release. Together with the three near-infrared Two Micron All-Survey (2MASS) filters \citep{2mass}, the peaks of the spectral energy distributions of any nearby brown dwarfs with $T_{eff} \gtrsim 300$K are well sampled \citep{Mainzer:2011ys}.  

Although hundreds of brown dwarfs with M, L, or T spectral types are now known, very few of them are closer than ten parsecs to the Sun. The nearest known L dwarfs with trigonometric parallaxes are
the L8 dwarf \object{DENIS-P J0255.0-4700} \citep{1999AJ....118.2466M} at $4.97 \pm 0.10$ pc \citep{2006AJ....132.1234C}, 
the L5 dwarf \object{2MASSW J1507476-162738} \citep{2000AJ....119..369R} at $7.33 \pm 0.03$ pc \citep{dahn}, 
the L5 dwarf \object{2MASSI J0835425-081923} \citep{2003AJ....126.2421C} at  $8.5 \pm 0.8$ pc \citep{2011AJ....141...54A}, 
the L3.5 \object{2MASSW J0036159+182110} \citep{2000AJ....119..369R} at $8.76 \pm 0.06$ pc \citep{dahn}, 
and the L5.5 dwarf \object{2MASS J17502484$-$0016151} \citep{Kendall:2007uq} at $9.2 \pm 0.2$ \citep{2011AJ....141...54A}.  
The lithium M9 brown dwarf \object{LP 944-20} \citep{Tinney:1998kx}, measured at $4.97 \pm 0.11$ pc \citep{Tinney:1996vn}, also deserves mention. 
Other apparently nearby L dwarfs still lack reliable trigonometric parallaxes. The ``blue" L6 dwarf \object{SDSS J141624.08+134826.7} \citep{Bowler:2010lr,2010AJ....139.1045S} is thought to lie at $8.4 \pm 1.9$ or $8.0 \pm  1.6$ on the basis of photometry, while \citet{2010A&A...510L...8S} argue for a parallax distance of $7.9 \pm 1.7$ pc  based on limited astrometry.  \citet{Looper:2008lr} list five additional L dwarfs with estimated distances between 9.4 and 9.9 pc.\footnote{We follow \citet{2008ApJ...674..451B} in viewing  \object{2MASS J11263991-5003550} as a blue L4.5 at 15 pc rather than a blue L9 at 8 pc \citep{2007MNRAS.378..901F}. \citet{2006AJ....132.2360H} have shown by trigonometric parallax that the \object{GJ 1001} (LHS 102) system \citep{EROS-Collaboration:1999vn} lies beyond 10 parsecs.}
Along with these warmer brown dwarfs, more than a dozen T dwarfs are known within ten parsecs. WISE, whose filters are ideally suited for finding cold late-type T dwarfs, is already adding to the census: \citet{Mainzer:2011ys} report a late-T dwarf at $d \approx 6-10$ pc, while \citet{Scholz:2011zr} have identified two high proper motion, late-T dwarfs with $d \approx 5$ pc using WISE, 2MASS and SDSS.  

L dwarfs are useful probes for studying atmospheric properties of objects below $2200$K, where clouds and dust opacity are important to the observed spectral energy distribution. 
In addition to their intrinsic interest, the latest-type L dwarfs are potential analogs to massive hot exoplanets such as those found around HR 8799 \citep{Bowler:2010lp,2010Natur.468.1080M,Barman:2011fk}.  Because they are isolated in the field rather than found orbiting a nearby bright star they can be studied and characterized in great detail making them vital test beds for refinement. 

We have initiated an effort to detect high proper motion stars by comparing the WISE and 2MASS surveys. Here, we report the identification of a bona-fide, nearby late-L brown dwarf, WISEP J180026.60+013453, as a result of this effort. We present our observations and data analysis in Section~\ref{da}. We discuss the possibility of binarity in Section~\ref{bin}.  

\section{Identification and Observations\label{da}}

\subsection{WISE Selection}

\begin{figure}
\plotone{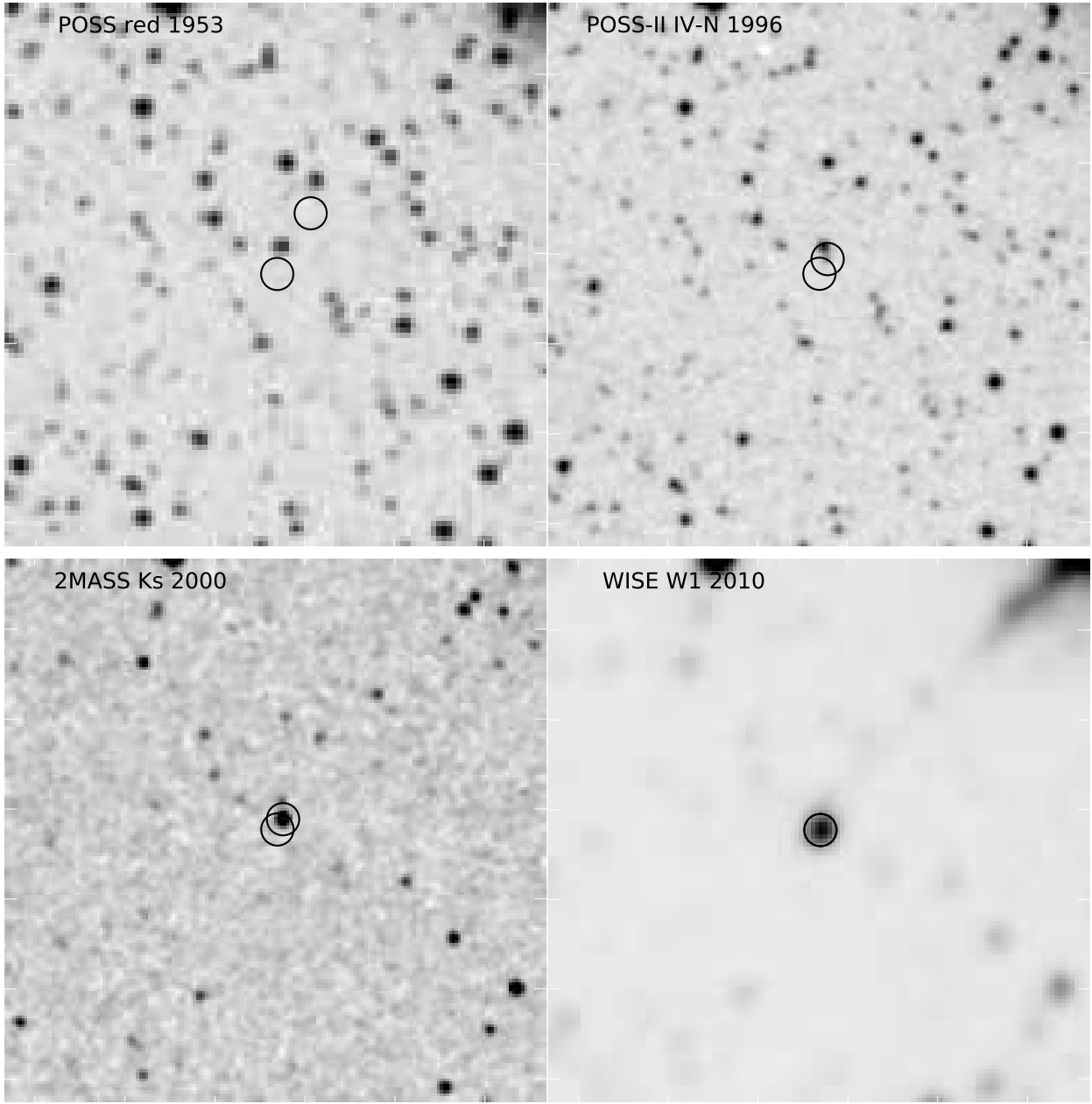}
\caption{Finder charts for WISEP J180026.60+013453.1: Upper left, July 1953 Palomar Sky Survey (POSS) red plate. Upper right, July 1996 Second Palomar Sky Survey (POSS-II) infrared plate. Lower left, September 2000 2MASS K$_s$ band. Lower right, March 2010 WISE W1 band.  In each case, the central circle marks the observed WISE position and an additional circle marks the expected position at that epoch. All are 3 arcminutes on a side.    
\label{fig1}}
\end{figure}

We selected WISE sources with $|b|>10$ that were detected at W1 ($3.4 \mu$m), W2 ($4.6 \mu$m) and W3 ($12\mu$m) with W1$<12$ but had no 2MASS counterpart within 3 arcseconds according to the WISE Preliminary Source Catalog.  All stars or warm brown dwarfs with W1$<12$ should have 2MASS detections at K$_s$ band. The WISE, 2MASS, and photographic (Digitized Sky Survey, DSS) images were examined by eye to look for proper motion candidates. Almost all stars have W1$-$W2$< 0.3$ and W1$-$W3$< 0.7$, but we noted a number of candidate proper motion objects with redder W1-W2 colors. Most of these turned out to be known late-M or L dwarfs, but the source WISEP J180026.60+013453.1 (Table~\ref{tab1}) was previously unrecognized. As seen in Figure~\ref{fig1}, this WISE source may be paired with 2MASS J18002648+0134565 (J$=14.30\pm 0.04$, H$=13.12 \pm 0.04$, K$_s=12.42\pm0.03$), which otherwise would not have a WISE counterpart. We hereafter abbreviate the source name as W1800+0134. The object was likely overlooked previously because it lies near the Galactic Plane ($b=12^\circ$), although the field is not very crowded. 

The apparent proper motion from 2MASS to WISE is $\mu_{\alpha} = 0.20 \pm 0.02$  arcsec yr$^{-1}$
and  $\mu_{\delta} = -0.36 \pm 0.02$ arcsec yr$^{-1}$. This is relative to other stars in the image, with uncertainties based on the  uncertainties of the WISE and 2MASS astrometry. However, because the observations were taken in 2010 March 20-23 (WISE) and 2000 September 23 (2MASS), parallax will affect the observed positions. Using the NOVAS V3.0 software  \citep{novas}, we find that the offset in RA is 230 mas if  W1800+0134 is 8.8 pc away, as estimated later.  Applying this correction, we find  $\mu_{RA} = 0.22 \pm 0.02$ arcsec yr$^{-1}$  and  $\mu_{DEC} = -0.36 \pm 0.02$ arcsec yr$^{-1}$.  A possible very faint source is seen at the predicted position on the infrared (I-band) Second Palomar Sky Survey plate \citep{Reid:1991rt}, but as expected for an L dwarf, it is not detected on the blue or red photographic plates (Figure~\ref{fig1}.)  We also identify W1800+0134 with DENIS J180026.4+013457, J$=14.07 \pm 0.10$ and K$=12.45 \pm 0.12$ in the 3rd release of the DENIS database. It was apparently not detected in the I band images, suggesting I$\gtrsim 18$. Because the DENIS observation is close in time to the 2MASS observation but is lower signal-to-noise, the detection confirms that W1800 has moved but was not included in the proper motion determination.  
 
\subsection{Near-Infrared Spectroscopy}

A low-resolution IRTF SpeX \citep{2003PASP..115..362R} spectrum was obtained on 2011 June 22 (UT) and processed using SpexTool \citep{2003PASP..115..389V,2004PASP..116..362C}.  
Spectra were obtained in prism mode using the 0.8'' slit aligned at the parallactic angle.  The resolution of the corresponding data spanning $0.7-2.5 \mu$m was $\lambda/\Delta \lambda \approx 90$.  Conditions were clear and the seeing was 0.7'' at K.  We obtained 4 individual exposure times of 120s in an ABBA dither pattern along the slit.  Immediately after the science observations we observed the A0V star \object{HD 171149}  at a similar airmass for telluric corrections and flux calibration.  Internal flat-field and Ar arc lamp exposures were acquired for pixel response and wavelength calibration, respectively.

\begin{figure}
\plotone{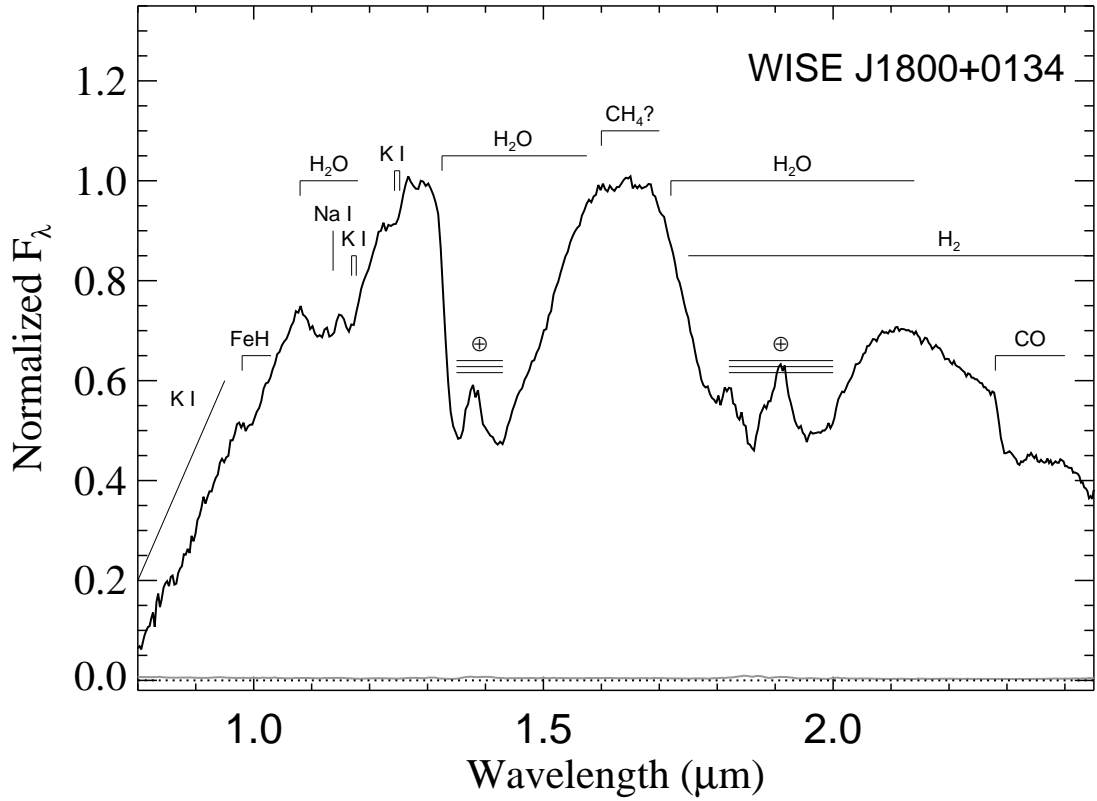}
\caption{Low-resolution near-infrared spectrum of W1800+0134 obtained with SpeX.  Data are normalized at 1.27~$\micron$,~ and the uncertainty spectrum is shown in gray.  Prominent atomic and molecular features are labeled. The CH$_4$ region at 1.6 \micron shows some flattening, but fits to the spectrum including an unresolved T dwarf companion are not statistically robust.
\label{fig2}}
\end{figure}

Our spectrum is shown in Figure~\ref{fig2}. We fit the spectrum to L and T dwarf templates also observed with SpeX.\footnote{Low-resolution L and T dwarf templates were drawn from the SpeX Prism Spectral Libraries, \url{http://www.browndwarfs.org/spexprism}.} The best matching  template is \object{2MASS J02572581-3105523}, classified as L8 in the optical by \citet{2008ApJ...689.1295K} on the \citet{1999ApJ...519..802K} system and observed with SpeX by \citet{Siegler:2007lr}.  The indices and spectral type relations defined by \citet{2007ApJ...659..655B} give L$6.5 \pm 0.07$ (H2O-J),  L$5.8\pm 0.07$ (H2O-H), L$6.9 \pm 0.09$ (H2O-K), and L$7.7 \pm 0.04$ (CH4-K), for a mean classification of L$7 \pm 1$.  The scatter is typical for near-infrared spectral typing of L dwarfs (e.g., \citealt{2007ApJ...659..655B}; \citealt{Stephens:2009qy}).  Combining the index type with the best-fit template type, we adopt a near-infrared spectral classification of L7.5$\pm$0.5 for W1800+0134.

Using the \citet{Looper:2008fk} spectral type-absolute magnitude relationships, and propagating uncertainties through Monte Carlo analysis, we estimate a distance of $8.8 \pm 1.0$ pc from the 2MASS J and K$_s$ photometry. Comparison of this estimated distance to the L dwarf distances listed in Section~\ref{intro} suggests W1800+0134 ranks as the sixth closest known L dwarf, but even a one parsec change could move it up to third or down to eighth or later.  In any case, a trigonometric parallax is needed to realize W1800+0134's scientific potential as one of the closest, brightest late-L dwarfs for detailed study.  

\subsection{Physical Properties}

The \citet{Looper:2008fk} polynomial fits to the \citet{2004AJ....127.3516G} temperature scale indicate that $T_{eff} = 1490$K and $BC_K = 3.24$  for L7.5 dwarfs. With our estimated distance, W1800+0134 is then $\log L/L_\odot \approx -4.5 \pm 0.3$.  The \citet{Stephens:2009qy} relations, also based on a reanalysis of  \citet{2004AJ....127.3516G}, make them somewhat cooler at $T_{eff} = 1430 \pm 100$K.   The tangential velocity ($18 \pm 2$ km s$^{-1}$) is below the median for late L dwarfs \citep{2009AJ....137....1F}. According to the \citet{1997ApJ...491..856B} models, for ages of 0.5, 1, and 5 Gyr, the mass of a $T_{eff} = 1430$K dwarf is $0.04 M_\odot$, $0.05 M_\odot$, and $0.074 M_\odot$ respectively; the \citet{2000ApJ...542..464C} models give similar results. For all ages, it is below the hydrogen-burning limit.  An optical spectrum could determine whether W1800+0134 is above or below the lithium-burning limit \citep{1993ApJ...404L..17M,2000AJ....120..447K}.  

\section{Is W1800+0134 a Binary? \label{bin}}

Some $\sim20\%$ of very-low-mass binaries in magnitude-limited samples are typically resolved as doubles \citep{2003AJ....125.3302G,2003AJ....126.1526B,2006ApJS..166..585B}.  If W1800+0134 were a near-equal luminosity double, the estimated distance would increase to $\sim 12$ pc.  The low-resolution spectrum shows some evidence of structure in the H-band around the region of the 1.6 $\micron$ CH$_4$ band, but no indication of CH$_4$ at K-band (see Figure~\ref{fig2}.) Unresolved L dwarf plus T dwarf binary spectra frequently show this pattern of CH$_4$ onset in the near-infrared \citep{Cruz:2004fr,2007ApJ...659..655B}. To investigate this possibility, we compared the SpeX spectrum to combinations of L and T dwarf binary templates from the SpeX Prism Spectral Libraries, following the procedures described in \citet{2007ApJ...659..655B}. The best fit binary has the same $\chi^2$ residual as the best fit single source; as such, there is no statistically robust evidence of W1800+0134 being an L plus T dwarf binary.

The 2MASS J-K$_s = 1.88\pm 0.05$ color is neither unusually blue nor red compared to other L7-L8 dwarfs \citep{2009AJ....137....1F,2010AJ....139.1808S}. The DENIS non-detection at I-band (corresponding to $I-J \gtrsim 3.7$) is also consistent with the observed ($I-J \approx 4$) colors of other L7-L8 dwarfs \citep{2008MNRAS.383..831P}. To assess whether the WISE colors are typical, we have searched for known objects with an L7, L7.5, or L8 classification in the WISE Preliminary database with uncertainties less than 0.1 in W1 and W2 and 0.2 in W3. The seven matching dwarfs are listed in Table~\ref{tab2}.  The observed mean and standard deviations of the apparently single objects are W1$-$W2 $= 0.54 \pm 0.05$ and W2$-$W3$ = 1.01 \pm 0.09$, consistent with W1800+0134's colors ($0.50 \pm 0.06$, $1.16 \pm 0.08$). The doubles \object{SDSSp J042348.57-041403.5} and \object{2MASS J05185995-2828372} are somewhat redder but the uncertainties in W$_3$ are large (0.08 and 0.17 respectively.)

We note for completeness that we found no other common proper motion companions within 30 arcminutes of W1800+0134, in WISE or 2MASS, ruling out widely-separated main sequence to late-T brown dwarf companions to a projected separation of 15,000 AU.

\section{Conclusion}

We have identified a previously overlooked bright and nearby L7.5 dwarf, near the Galactic Plane. Together with the recent discoveries of the M9 \object{DENIS-P J115927.4-524718} at 9.6 pc ($b = 9^\circ$; \citealt{2008MNRAS.383..831P}), the T6 \object{DENIS J081730.0-615520} at 4.9 pc ($b = -14^\circ$; \citealt{2010ApJ...718L..38A}), the T9 \object{UGPS J072227.51-054031.2} at 4.1 pc ($b = 4^\circ$; \citealt{2010MNRAS.408L..56L}), and the T9 \object{WISEPC J045853.90+643451.9} at 6-10 pc ($b=13^\circ$; \citealt{Mainzer:2011ys}), it is clear that multi-color, multi-survey searches for nearby brown dwarfs in the Galactic Plane region are now feasible.  Such objects are important because their proximity enables more detailed studies of the physical properties of brown dwarfs, as well as any potential planetary companions.  As a very bright late-L dwarf with normal colors, studies of W1800+0134's cloud properties through variability and polarization measurements (e.g., \citealt{Artigau:2009ly, Goldman:2009mz}), and rotation through high-resolution spectroscopy (e.g., \citealt{2006ApJ...647.1405Z}), will be helpful in characterizing the structure and dynamics of cloudy atmospheres at the L/T transition, and will be relevant to exploring the cloud properties of hot exoplanets \citep{2011arXiv1102.5089M}.

\acknowledgments

We thank the referee (Chris Tinney) for useful comments.  
This publication makes use of data products from the Wide-field Infrared Survey Explorer, which is a joint project of the University of California, Los Angeles, and the Jet Propulsion Laboratory/California Institute of Technology, funded by the National Aeronautics and Space Administration. This publication makes use of data products from the Two Micron All Sky Survey, which is a joint project of the University of Massachusetts and the Infrared Processing and Analysis Center/California Institute of Technology, funded by the National Aeronautics and Space Administration and the National Science Foundation. This research has made use of the NASA/ IPAC Infrared Science Archive, which is operated by the Jet Propulsion Laboratory, California Institute of Technology, under contract with NASA. This research has made use of the VizieR catalogue access tool, CDS, Strasbourg, France. This research has made use of the SIMBAD database, operated at CDS, Strasbourg, France. The Digitized Sky Surveys were produced at the Space Telescope Science Institute under U.S. Government grant NAG W-2166. The images of these surveys are based on photographic data obtained using the Oschin Schmidt Telescope on Palomar Mountain and the UK Schmidt Telescope. The plates were processed into the present compressed digital form with the permission of these institutions. This research made use of APLpy, an open-source plotting package for Python hosted at \url{http://aplpy.github.com}. This research has benefitted from the M, L, and T dwarf compendium housed at \url{DwarfArchives.org} and maintained by Chris Gelino, Davy Kirkpatrick, and Adam Burgasser. This research has benefitted from the SpeX Prism Spectral Libraries, maintained by Adam Burgasser at \url{http://www.browndwarfs.org/spexprism}






\begin{deluxetable}{lc}
\tablewidth{0pc}
\tabletypesize{\footnotesize}
\tablenum{1} \label{tab1}
\tablecaption{WISEP J180026.60+013453.1}
\tablehead{
\colhead{Parameter} & 
\colhead{W1800+0134} }
\startdata
WISE RA(J2000) & 18 00 26.60 \\
WISE Dec (J2000) & +01 34 53.1\\
WISE Epoch & 2010.22 \\
2MASS J [mag] &  $14.30 \pm 0.04$ \\
2MASS H [mag] & $13.12 \pm 0.04$ \\
2MASS K$_s$ [mag] & $12.42 \pm 0.03$ \\
WISE W1 [mag] & $11.53 \pm 0.05$ \\ 
WISE W2 [mag] & $11.03 \pm 0.04$ \\
WISE W3 [mag] & $9.88 \pm 0.07$ \\
WISE W4 [mag] & $>8.39$ \\
$\mu_\alpha \cos \delta$ (mas/yr) & $220 \pm 20$ \\
$\mu_\delta$ (mas/yr) & $-360 \pm 20$ \\
Sp Type (Near-IR) & L$7.5 \pm 0.5$\\
distance (pc) & $8.8 \pm 1.0$ \\
$v_{tan}$ km s$^{-1}$ & $18 \pm 2$ \\
$T_{eff} (K)$            & $1430 \pm 100$ \\
$\log L/L_\odot$   & $-4.5 \pm 0.3$ \\
\enddata
\end{deluxetable}

\begin{deluxetable}{llrcccl}
\tablewidth{0pc}
\tabletypesize{\footnotesize}
\tablenum{2} \label{tab2}
\tablecaption{WISE Data for L7-L8 dwarfs}
\tablehead{
\colhead{Discovery Name} & 
\colhead{WISEP} & 
\colhead{W1} & 
\colhead{W1$-$W2} & 
\colhead{W2$-$W3} &
\colhead{Double?} &
\colhead{Ref.} }
\startdata 
\object{2MASS J03185403-3421292} & J031854.37$-$342128.7 & 12.62 & 0.50 & $0.97 \pm 0.11$ &\nodata  & 1 \\
\object{SDSSp J042348.57-041403.5} & J042348.32$-$041402.5 & 12.18 & 0.61 & $1.03 \pm 0.09$ & Y &2, 3 \\
\object{2MASS J05185995-2828372}  & J051859.89$-$282840.2 & 13.41 & 0.59 & $1.09 \pm 0.17$ & Y & 4, 3 \\
\object{2MASSI J0825196+211552}  & J082519.24+211548.5 & 12.08 & 0.52 & $1.02 \pm 0.09$ & \nodata  &5  \\
\object{SDSSp J085758.45+570851.4}  &J085757.95+570847.5 & 12.03 & 0.61 & $1.12 \pm 0.06$ & \nodata & 2 \\
\object{2MASSW J1632291+190441} & J163229.39+190439.9 & 13.14 & 0.52 & $0.92 \pm 0.19$ & \nodata &  6 \\
\object{2MASSW J1728114+394859} & J172811.53+394859.1 & 13.10 &  0.49 & $0.94 \pm 0.12$  & Y & 5, 7 \\
WISEP J180026.60+013453.1 & J180026.60+013453.1& 11.53 & 0.55 & $1.16 \pm 0.08$ & \nodata & 8 \\
\enddata 
\tablecomments{The uncertainty in W$_1$ is 0.03 mags and in W$_1$-W$_2$ is 0.04 mags for all objects.}
\tablerefs{1. \citet{2008ApJ...689.1295K} 2. \citet{2002ApJ...564..466G} 3. \citet{2006ApJS..166..585B} 4. \citet{Cruz:2004fr} 5. \citet{2000AJ....120..447K}  6. \citet{1999ApJ...519..802K}  7. \citet{2003AJ....125.3302G} 8. This paper }
\end{deluxetable}

\end{document}